\documentclass[sigconf]{acmart}
\usepackage{bm}
\usepackage{subfigure}
\newtheorem{defi}{Definition}
\newtheorem{prop}{Proposition}
\newtheorem{lem}{Lemma}

\newtheorem{ass}{Assumption}
\AtBeginDocument{%
  }

\copyrightyear{2023}
\acmYear{2023}
\setcopyright{rightsretained}
\acmConference[e-Energy '23]{The 14th ACM International Conference on Future Energy Systems}{June 20--23, 2023}{Orlando, FL, USA}
\acmBooktitle{The 14th ACM International Conference on Future Energy Systems (e-Energy '23), June 20--23, 2023, Orlando, FL, USA}
\acmDOI{10.1145/3575813.3595201}
\acmISBN{979-8-4007-0032-3/23/06}

\begin{document}

\title{Uniform Pricing vs Pay as Bid in 100\%-Renewables Electricity Markets: A Game-theoretical Analysis}
\author{Dongwei Zhao }
%\authornote{Both authors contributed equally to this research.}
\email{zhaodw@mit.edu}
%\orcid{1234-5678-9012}
\authornotemark[1]
\affiliation{%
  \institution{Massachusetts Institute of Technology}
  \streetaddress{P.O. Box 1212}
  \city{Cambridge}
  \state{MA}
  \country{USA}
  \postcode{43017-6221}
}

\author{Audun Botterud}
\email{audunb@mit.edu}
\affiliation{%
  \institution{Massachusetts Institute of Technology}
  \city{Cambridge}
  \state{MA}
  \country{USA}}

\author{Marija Ilic}
\email{ilic@mit.edu}
\affiliation{%
  \institution{Massachusetts Institute of Technology}
  \city{Cambridge}
  \state{MA}
  \country{USA}}

% \author{Ben Trovato}
% \authornote{Both authors contributed equally to this research.}
% \email{trovato@corporation.com}
% \orcid{1234-5678-9012}
% \author{G.K.M. Tobin}
% \authornotemark[1]
% \email{webmaster@marysville-ohio.com}
% \affiliation{%
%   \institution{Institute for Clarity in Documentation}
%   \streetaddress{P.O. Box 1212}
%   \city{Dublin}
%   \state{Ohio}
%   \country{USA}
%   \postcode{43017-6221}
% }

% \author{Lars Th{\o}rv{\"a}ld}
% \affiliation{%
%   \institution{The Th{\o}rv{\"a}ld Group}
%   \streetaddress{1 Th{\o}rv{\"a}ld Circle}
%   \city{Hekla}
%   \country{Iceland}}
% \email{larst@affiliation.org}

% \author{Valerie B\'eranger}
% \affiliation{%
%   \institution{Inria Paris-Rocquencourt}
%   \city{Rocquencourt}
%   \country{France}
% }

% \author{Aparna Patel}
% \affiliation{%
%  \institution{Rajiv Gandhi University}
%  \streetaddress{Rono-Hills}
%  \city{Doimukh}
%  \state{Arunachal Pradesh}
%  \country{India}}

% \author{Huifen Chan}
% \affiliation{%
%   \institution{Tsinghua University}
%   \streetaddress{30 Shuangqing Rd}
%   \city{Haidian Qu}
%   \state{Beijing Shi}
%   \country{China}}

% \author{Charles Palmer}
% \affiliation{%
%   \institution{Palmer Research Laboratories}
%   \streetaddress{8600 Datapoint Drive}
%   \city{San Antonio}
%   \state{Texas}
%   \country{USA}
%   \postcode{78229}}
% \email{cpalmer@prl.com}

%\renewcommand{\shortauthors}{Trovato et al.}

\begin{abstract}
	
This paper evaluates market equilibrium  under different pricing mechanisms in a two-settlement  100\%-renewables  electricity market. Given general probability distributions of renewable energy,   we establish game-theoretical models to analyze equilibrium bidding strategies,  market prices, and profits under uniform pricing (\textbf{UP}) and pay-as-bid pricing (\textbf{PAB}). We prove that \textbf{UP} can incentivize suppliers to withhold bidding quantities and lead to price spikes. \textbf{PAB} can reduce the market price, but it may lead to a mixed-strategy price equilibrium.  Then, we present a regulated uniform pricing scheme (\textbf{RUP}) based on suppliers' marginal costs that include penalty costs for real-time deviations. We show that \textbf{RUP}  can achieve lower yet positive prices and profits  compared with  \textbf{PAB} in a duopoly market, which approximates the least-cost system outcome. Simulations with  synthetic and real data  find that under \textbf{PAB} and \textbf{RUP}, higher uncertainty of renewables and real-time shortage penalty prices can increase the market price by encouraging lower bidding quantities, thereby increasing suppliers' profits.
  
\end{abstract}




\maketitle

\section{Introduction}
Uniform pricing (\textbf{UP}) is a widely adopted auction design in deregulated electricity markets. The clearing market price under  \textbf{UP} is identical to all the suppliers (in one location or zone), regardless of their bidding offers \cite{fundamentals}. If there exists no market power, \textbf{UP} can achieve market efficiency \cite{schweppe2013spot}, and is also transparent in terms of selecting the least-cost suppliers \cite{lin2017electricity}. In practice,  suppliers can exercise market power, such as reporting a price higher than its marginal cost or withholding capacity, to increase the market price \cite{kahn2001uniform}\cite{zhao2022market}. For conventional dispatchable generators, it may not be difficult for the market monitor to mitigate market power by comparing offers to marginal costs and generation capacities \cite{graf2021market}.  

However, the \textbf{UP} mechanism still faces some challenges when variable renewable energy (VRE) dominates the market. First, VRE has zero marginal costs, which may cause low and volatile day-ahead market prices if suppliers bid at true marginal costs \cite{leslie2020designing}.  Second, VRE generation is variable and uncertain. Re-balancing due to uncertain outputs may cause substantial real-time costs. The above two considerations may aggravate profit losses to suppliers and encourage strategic behaviors, e.g.,  withholding bidding quantities, to increase market prices. There are discussions around allowing VRE suppliers to offer at prices higher than zero marginal costs \cite{shen2022valuing}. However, since the re-balancing cost and generation capacity of VRE are uncertain and variable,  it is challenging for the market monitor to assess what constitutes competitive bidding.

The challenges of implementing \textbf{UP} under high VRE levels open doors for discussing other auction designs, such as pay-as-bid pricing (\textbf{PAB}), where suppliers are paid directly at bidding prices. By introducing price competition, \textbf{PAB} may reduce the average market prices and revenues compared with  \textbf{UP}  and thus benefit consumers \cite{son2004short}\cite{fabra2006designing}\cite{bashi2018comparative}. One drawback of \textbf{PAB} is that it may fail to reveal the true marginal costs of suppliers and thus leads to market inefficiency \cite{akbari2020electricity}. The market price under \textbf{PAB} is also hard to predict and regulate \cite{heim2021pay}. However, since VRE suppliers' re-balancing cost and generation capacity are uncertain, \textbf{UP} may also face similar challenges to \textbf{PAB} under high penetration levels of VRE.

Electricity market design for high penetration levels of VRE is still an open question, and the literature on equilibrium analysis under \textbf{PAB} and \textbf{UP} is under-explored. Son \textit{et al.} \cite{son2004short} analyzed the Nash equilibrium of suppliers' bidding strategies based on game theory but did not take into account uncertainty from VRE. In contrast, some works analyzed the Nash equilibrium of VRE suppliers' bidding strategies. Ju \textit{et al.} \cite{ju2022distribution} and Taylor \textit{et al.}  \cite{taylor2016strategic} focused on \textbf{UP}  and \textbf{PAB}, respectively,  which both only focused on the single-settlement market and neglected uncertainty-related costs in real time. Zhao \textit{et al.} \cite{zhao2019storage} analyzed the market equilibrium under \textbf{PAB} for VRE considering real-time penalty costs. However,  the work did not characterize the equilibrium under \textbf{UP} nor make a comparison between  \textbf{UP}  and \textbf{PAB}. In contrast, in this paper, we provide an analysis of market equilibrium under \textbf{UP} and compare it with  \textbf{PAB}. We also propose a regulated uniform pricing scheme (\textbf{RUP}) that takes into account real-time deviation costs. We summarize the  contributions of this paper in the following.

\textit{Nash equilibrium analysis in a two-settlement market with VRE only:} We evaluate the strategic behaviors of suppliers in VRE-only markets, which takes account of day-ahead revenues and real-time shortage penalty costs. We establish game-theoretical frameworks to model suppliers' bidding strategies under \textbf{UP} and \textbf{PAB}. Given general probability distributions of VRE, we characterize the Nash equilibrium of suppliers' bidding prices and quantities, and analyze equilibrium market prices and  profits. 

\textit{Market mechanism comparison:} We compare  equilibrium prices and profits under \textbf{UP} and  \textbf{PAB}. We prove that under \textbf{UP} suppliers can exercise market power to cause price spikes. \textbf{PAB} can reduce the market price compared with \textbf{UP} by differentiating prices. However,  no pure strategy equilibrium exists under \textbf{PAB}. 
We propose \textbf{RUP} as a potential regulation benchmark for renewable energy,  which can achieve lower yet positive prices and profits compared with \textbf{PAB} under duopoly competition.

\textit{Simulation insights}: We utilize both synthetic- and real-world data to perform simulations, which show that  \textbf{RUP} leads to the lowest market prices and profits. We also observe that under \textbf{PAB} and \textbf{RUP}, higher uncertainty in VRE and a higher real-time penalty price may increase  market prices by encouraging conservative bidding quantities thereby increasing suppliers' profits.

%The rest of the paper is organized as follows. In Section \ref{sec:sys}, we introduce the general system model. In Sections \ref{sec:up}-\ref{sec:sup}, we show the analysis for \textbf{UP}, \textbf{PAB}, and \textbf{RUP}, respectively.  We perform simulations based on synthetic and real-world data in Section \ref{sec:sim} and conclude in Section \ref{sec:con}.

\section{System model}\label{sec:sys}

We introduce the model of renewable-energy suppliers, the setting of electricity markets, and some assumptions.

We consider a 100\%-VRE electricity market (e.g., solar and wind), where a set of VRE suppliers is denoted by $\mathcal{I}=\{1,2,...,I\}$. In the rest of the paper, we  use suppliers to refer to VRE suppliers.
For  one certain hour, the  output  of  supplier $i\in \mathcal {I}$ is denoted as a random variable $X_i$, which has the support over  $[0,\overline{X}_i]$. We assume that the random generation $X_i$ has a  continuous cumulative distribution function (CDF) $F_i$ with the probability  density  function (PDF) $f_i$. We assume zero marginal production costs for VRE suppliers.

We model a two-settlement electricity market, which consists of a day-ahead  market (DAM) and a real-time  market (RTM)  \cite{bringwind}. 
In the DAM, any supplier $i$ submits the bidding price $p_i$ with a cap $\bar{p}$, quantity $q_i$, or supply curve $Q_i(\cdot)$ to the system operator. Based on suppliers' bidding strategies and demand $D$, the system operator clears the market with the price $\pi_i$ and supplier delivery commitment $\bm{x}=(x_i,i\in \mathcal{I})$. In the RTM, if supplier $i$'s actual generation falls short of the committed quantity, i.e., $x_i>X_i$, it needs to pay the real-time penalty  price $\lambda$, resulting in the penalty cost $\lambda( x_i-X_i)$.  We assume that $\lambda>0$ is fixed (or the expected value of the random penalty price independent of random generations). Such a penalty cost will reflect the cost of other flexible resources addressing the deviation. Overall, the profit of supplier $i$ is
 	\begin{align}
 	R_i=&x_i \cdot \pi_i -\mathbb{E}_{X_i}\left[\lambda\cdot  (x_i-X_i)^+\right].\label{eq:profitx}
 \end{align}	

 We clarify some  assumptions in this paper which we will further generalize in  future work.
\begin{ass}
	(i)	The system demand  $D>0$ is fixed and inelastic;  (ii) The  penalty price is positive, i.e., $\lambda>0$;
 (iii) There is no reward or penalty on the excessive generation (i.e., $x_i<X_i$) in real time, and the excessive generations are simply curtailed; (iv) Suppliers are price-takers in the RTM;  (v) Suppliers have complete market information.
\end{ass}

In Sections \ref{sec:up}- \ref{sec:sup}, we will discuss different market clearing mechanisms in the day-ahead market: \textbf{UP}, \textbf{PAB}, and \textbf{RUP}, respectively. We will analyze the equilibrium market  price and profits based on game-theoretical models.

\section{Uniform pricing}\label{sec:up}

We will characterize the Nash equilibrium of suppliers' bidding prices and quantities under \textbf{UP}. 

Before going into details, we will first present an optimal day-ahead commitment for suppliers based on the cleared price, which can help analyze bidding quantities under different mechanisms later. Based on the profit formulation \eqref{eq:profitx},  Lemma \ref{lem:quantity}  characterizes the optimal day-ahead committed quantity.

\begin{lem}[optimal commitment]\label{lem:quantity}
If supplier $i$ is paid at the price $\pi_i$ in the DAM, the cleared commitment	$x_i=y_i^*(\pi_i)$ in the following  will maximize its profit.
\vspace{-0.5ex}
	\begin{align}
		y_i^*(\pi_i)=	F_i^{-1}\left(\min(\frac{\pi_i}{\lambda},1)\right).
	\end{align}
	%Furthermore, supplier $i$' s profit $R_i$ increases over $x \in [0,	y_i^*(\pi_i)]$ and decreases over $x \in [	y_i^*(\pi_i),\bar{p}]$.
\end{lem}
Lemma 1 is easily proved based on the first order condition of \eqref{eq:profitx}. Here	$y_i^*(\pi_i)$ is non-decreasing over $\pi_i\in [0,\bar{p}]$.  When the price is zero, any supplier should commit zero quantity in the DAM and zero profits, i.e., $y_i^*(0)=0$ to avoid penalty costs in real time. If $\pi_i\geq \lambda$, the supplier will just bid the maximum quantity $\overline{X}_i$. Next, under \textbf{UP},  we will first consider that  suppliers are required to bid zero prices and then generalize it to any prices in Appendix B.

\subsection{Pricing mechanism} Each supplier bids the price $p_i=0$ and quantity $q_i\geq0$. The clearing  price $\pi^*$ and commitment $\bm{x}^*$ are characterized  in the following.\footnote{Since all suppliers bid the same price, we assume that the system operator will allocate demand by a random  merit order.}
\vspace{-1ex}
\begin{subequations}\label{eq:upzero}
	\begin{align}
		&\text{If~} \sum_i q_i\leq  D,  \pi^*=\bar{p} \text{~and~} x_i^*=q_i.\\
		&	\text{If~} \sum_i q_i> D,  \pi^*=0 \text{~and~}  \sum_{i\in \mathcal{I}} x_i^*=D.
	\end{align}
	\end{subequations}

This mechanism leads to bipolar prices. Note that at the point $\sum_i q_i= D$, the market price is not continuous. To maintain a pure strategy Nash equilibrium, we consider that the market price is right-continuous in $\sum_i q_i$, i.e.,  	 $\pi=\bar{p}$ when $\sum_iq_i=D$. 

\subsection{Game-theoretical model} Each supplier $i$ decides on the bidding quantity $q_i\geq0$ to maximize its profit. We formulate a game-theoretical model as suppliers' decisions are coupled due to the market clearing price. 

\begin{itemize}
	\item Players: Suppliers $\mathcal{I}$
	\item Strategy: Bidding quantity $0\leq q_i\leq \overline{X}_i$ of supplier $i$
	\item Payoff: Profit $R_i$ of supplier $i$
 \vspace{-0.5ex}
	\begin{align}
		R_i(q_i,q_{-i})=x_i^* (\bm{q}) \cdot \pi^*(\bm{q}) -\mathbb{E}_{X_i}\left[\lambda\cdot  (x_i^*(\bm{q})-X_i)^+\right], 
	\end{align}	
where $\pi^*$  and $\bm{x}^*$ are given by \eqref{eq:upzero}.

\end{itemize}

Definition  \ref{def:pureprice} defines the pure-strategy bidding-quantity equilibrium of suppliers,  where no supplier can increase its profit through unilateral  deviation.  
\begin{defi}[pure quantity equilibrium]\label{def:pureprice}
A bidding quantity vector $\bm{q}^*$ is a pure price equilibrium if for any supplier $i\in \mathcal{I}$,
	\begin{align}
		&R_i\left({q}_i^*,{q}_{-i}^*\right)\geq R_i\left({q}_i,{q}_{-i}^*\right),~\forall q_i\geq 0.
	\end{align}
\end{defi}

\subsection{Nash equilibrium} 
The Nash equilibrium of bidding quantities is affected by supply and demand. First, if there is supply shortage, i.e., $\sum_i y_i^*({\bar{p}})\leq D$, any supplier $i$ just bids the quantity at $ y_i^*({\bar{p}})$, which leads to the price cap $\bar{p}$. It means there is not enough generation capacity in the market. However, if there is adequate supply, i.e., $\sum_i y_i^*({\bar{p}})> D$, the Nash equilibrium is given as follows.

\begin{prop}[Equilibrium bidding quantity $\bm{q}^*$]\label{prop:upq}
	  If 	$\sum_i y_i^*(\bar{p})> D$,  the following conditions give a Nash equilibrium.
		\begin{align}
			\text{(i)} ~\sum_i q_i^*= D;~ \text{(ii)}~ 0\leq q_i^*\leq  y_i^*(\bar{p}), \forall i\in \mathcal{I},\label{eq:upzeronash}
\end{align}
	where the market clearing price is $\pi^{*}=\bar{p}$. Furthermore, all the Nash equilibria satisfy $\pi^*=\bar{p}$ and $\sum_i q_i^*= D$.
\end{prop}

We show the proof in Appendix \ref{app:prop1}. Proposition \ref{prop:upq} shows that  the total bidding quantity of suppliers is exactly at the demand $D$, which achieves the price cap $\bar{p}$. Excessive bidding leads to zero price and deficit bidding will encourage some suppliers to bid more.  Since  suppliers bid the same zero price, it is not unique how to determine the merit order. Theoretically, some suppliers can get a larger share of demand while some get a much smaller one.

Although most  current markets require suppliers to bid at marginal costs, they cannot fully capture the cost incurred by renewables' uncertainty in  real time. Therefore, we further generalize the setting in Appendix \ref{app:generalize} and allow suppliers to bid any price instead of zero. One Nash equilibrium in the generalized case will coincide with Proposition \ref{prop:upq} at bidding price zero $p_i^*=0,\forall i$, where the clearing price reaches the price cap $\bar{p}$. 

%This Nash equilibrium, however, may not be unique. We give another example of Nash equilibrium in Appendix B, where the clearing market price may not be at $\bar{p}$ but still be manipulated by marginal suppliers whose bidding price will be the clearing  price.

In summary, under \textbf{UP}, suppliers can easily exercise market power to incur high prices. Since  suppliers have zero marginal costs with uncertain generations, it can be harder for the market monitor to regulate compared to conventional  generators.

\section{Pay-as-bid pricing}  \label{sec:pab}

Next, we will introduce the \textbf{PAB} mechanism, corresponding game-theoretical model, and analysis of Nash equilibrium.

 Under \textbf{PAB}, suppliers are paid at their bidding prices.
Each supplier $i$ decides the bidding price $0\leq p_i\leq \bar{p}$ and bidding quantity $0\leq q_i$ to maximize its profit $R_i$.  %following the solution to Problem \textbf{MO}. 
	\begin{align}
		R_i\left(\bm{p},\bm{q}\right)=&x^*(\bm{p},\bm{q}) \cdot  p_i - \mathbb{E}_{X_i}\left[\lambda\cdot  (x^*(\bm{p},\bm{q}) -X_i)^+\right],
	\end{align}	
where the commitment $\bm{x}^*$ follows the merit order given by the solution to Problem \textbf{MO} in Appendix B.

For the Nash equilibrium, first, if there is supply shortage $\sum_i y_i^*({\bar{p}})\leq D$, any supplier $i$ just bids the quantity at $ y_i^*({\bar{p}})$ and  the price cap $\bar{p}$. Then, if  $\sum_i y_i^*({\bar{p}})> D$, the Nash equilibrium results have been discussed in \cite{zhao2019storage}. In summary, for the bidding quantity, there is a weakly dominant strategy $q_i^*=\bm{y}^*({p}_i)$ for  supplier $i$ given its bidding price. For the price equilibrium, however, there is no pure-strategy price equilibrium,\footnote{If we generalize the  support from $[0,\overline{X}_i]$ to $[\underline{X}_i,\overline{X}_i]$ with $\underline{X}_i>0$, the  pure-strategy price equilibrium may exist due to the stable minimum generation $\underline{X}_i$. We will include this generalization in future work.}  but a mixed price equilibrium exists.

%\begin{defi}[mixed price equilibrium]\label{def:mix}
%	A vector of probability measures $\bm{\mu}^*$ is a mixed price equilibrium if, for any $i$, 
%	\begin{align*}
%		&\int_{{[0,\bar{p}]}^{I}} \pi_i^R \left(p_i,  {x}_i^*\left((p_i, {p}_{-i}),\bm{y}^*(p_i, p_{-i})\right) \right)d \left(\mu_i^{*}(p_i) \times{\mu}_{-i}^{*}({p}_{-i}) \right)\\
%		\geq &\int_{{[0,\bar{p}]}^{I}}\hspace{-0.6mm}	\pi_i^R \left(p_i,  {x}_i^*((p_i, {p}_{-i}),\bm{y}^*(p_i, p_{-i}))\right) d \left(\mu_i(p_i) \times {\mu}_{-i}^{*}({p}_{-i}) \right),
%	\end{align*}
%	for any measure  $\mu_i$.
%\end{defi}

The equilibrium price under \textbf{PAB}  is lower than the price cap $\bar{p}$. However, it is challenging to characterize or predict in practice the mixed price equilibrium under general probability distributions of renewables.  Also, a finite number of suppliers may still have market power to set a high price. Next, we examine a regulated supply-curve-based uniform pricing (\textbf{RUP}), which can further reduce the market price and provide benefits to consumers.

\section{Supply-curve-based uniform pricing} \label{sec:sup}

Under \textbf{RUP}, any supplier $i$ truthfully reports its inverse CDF function {$F_i^{-1}(\cdot)$}  of random generations, based on which we have  $y_i^*(\cdot)$ as the supply curve.  The system operator sets up the cumulative supply curve $Q(p)$.
\vspace{-2ex}
	\begin{align}
		Q(p)=\sum_{i \in \mathcal{I} } y_i^*({p}), ~p\leq \bar{p}.
	\end{align}
The operator clears the market with price $\pi^*$: If 	$Q(\bar{p})\geq D$, then $\pi^*$ is achieved at  $Q(\pi^*)=D$.  If 	$Q(\bar{p})< D$, then $\pi^*$ is achieved at 	$\bar{p}$ and the capacity is not adequate. 

We build the connection between  \textbf{RUP} and a benchmark of near-least system cost in Proposition \ref{prop:social}.
\begin{prop}\label{prop:social}
    The cleared day-ahead commitment  and market price under \textbf{RUP} give the optimal primal solution $\bm{x}^*$ and dual solution $\pi^e$, respectively, to the following problem. 
\begin{subequations}\label{eq:social}
	\begin{align}
 \min_{\bm{x}} &\sum_{i\in\mathcal{I}} \mathbb{E}_{X_i}\left[\lambda\cdot  (x_i-X_i)^+\right]+\bar{p}\cdot x_0\label{eq:socialobj}\\
 &\sum_{i\in\mathcal{I}} x_i+x_0=D,\hspace{25ex}:\pi^e\\
 &x_0\geq 0, x_i\geq0, \forall i \in \mathcal{I},
	\end{align}  
\end{subequations}
where $x_0$ denotes the lost load.
\end{prop}

The above proposition is easily proven based on KKT conditions. If suppliers' generation variables are independent, the objective function \eqref{eq:socialobj} is exactly the system cost. Since  suppliers make decisions based on their own generations, the generation correlation between suppliers cannot  be captured. Future work will consider how to distribute the correlation information among suppliers.

This mechanism captures the total marginal cost across the  DAM and RTM, making it suitable as a regulation benchmark. First, it approximates the least-cost system solution.  Second,  each supplier's profit is maximized at the clearing price.  Third, the clearing price is easily obtained compared with the mixed price equilibrium in \textbf{PAB}. Lastly, it will  lead to lower  prices compared with  \textbf{UP} and \textbf{PAB}. If we consider a duopoly case, the  price	under \textbf{RUP} (denoted by $\pi^{\text{RUP}}$) is lower than the  minimum expected equilibrium price  of two suppliers ($\pi^{\text{PAB}}$) under  \textbf{PAB}. We compare the prices in the following proposition, which is proved  in Appendix \ref{app:proof3}.

\begin{prop}[Equilibrium price comparison]\label{prop:pricecomp}
	Considering a duopoly market, the equilibrium (expected) market price satisfies
	\begin{align}
	\bar{p} \geq \pi^{\text{PAB}}\geq \pi^{\text{RUP}}>0. \label{eq:pricecomp}
	\end{align}
\end{prop}

If suppliers can report any supply curve, the Nash equilibrium in Proposition \ref{prop:uppq1} under uniform pricing is one possible result. The system operator needs to monitor and evaluate the generation information of generators to mitigate market power.

\section{Simulation results}	\label{sec:sim}

We simulate the results of market prices and suppliers' profits under \textbf{UP}, \textbf{PAB}, and \textbf{RUP} of two suppliers. The results show that \textbf{RUP} leads to the lowest price and profits for suppliers. We investigate the impact of generation uncertainty and  real-time penalty price based on synthetic data and real data, respectively.

We simulate two case studies: (i) we assume a truncated normal distribution of suppliers' renewable-energy generations and examine how the market price and suppliers' profits will change with generation uncertainty ; (ii) we use historical real data to establish the probability distribution of generations and investigate the impact of real-time penalty prices. We set demand at 2MW and set the bidding price cap at $\bar{p}=1$k\$/MWh.

\begin{figure}[t]
	\centering
	\hspace{-6ex}
	\subfigure[]{
		\raisebox{-1mm}	{\includegraphics[width=1.65in]{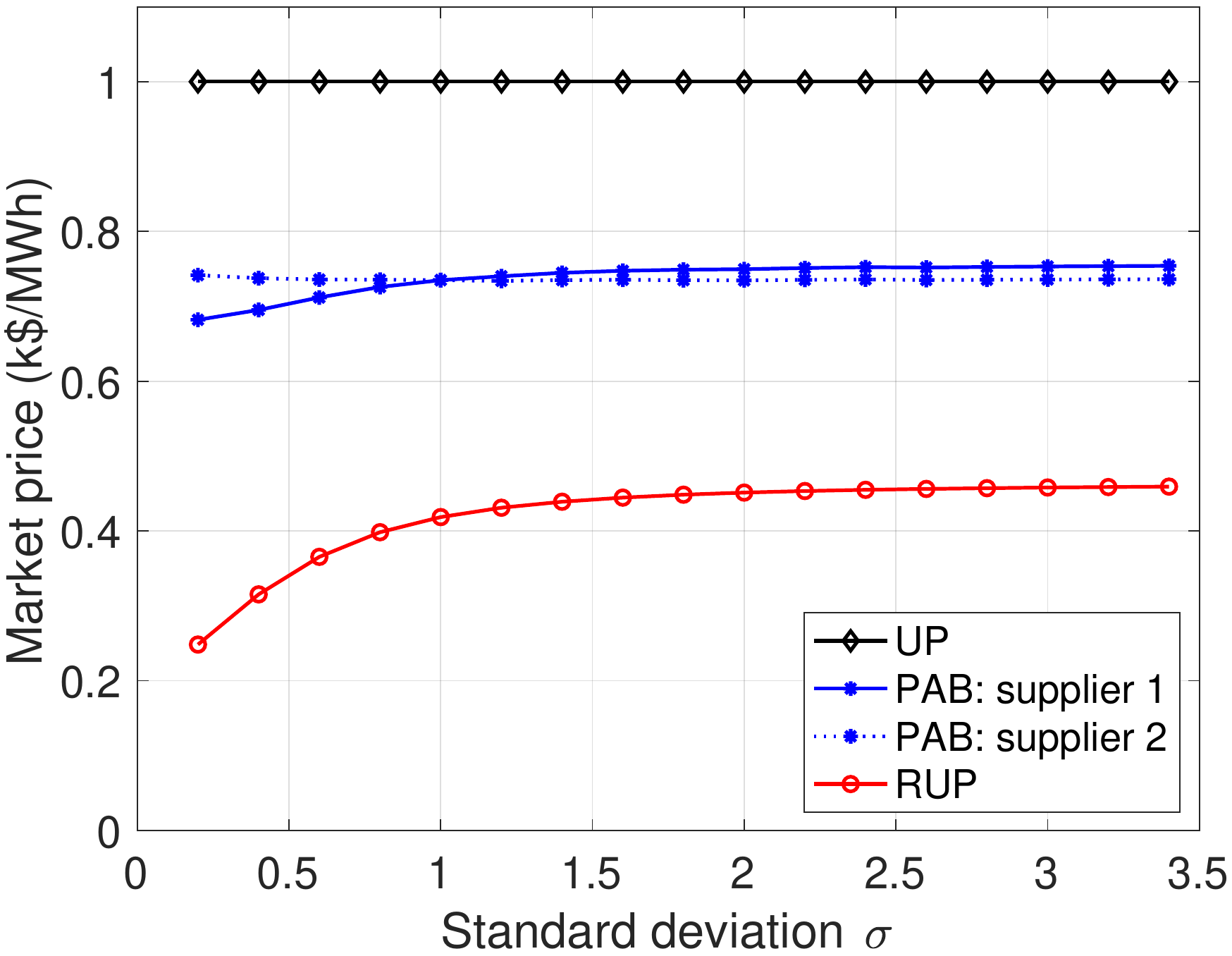}}}
	\hspace{-1ex}
	\subfigure[]{
		\raisebox{-1mm}		{\includegraphics[width=1.65in]{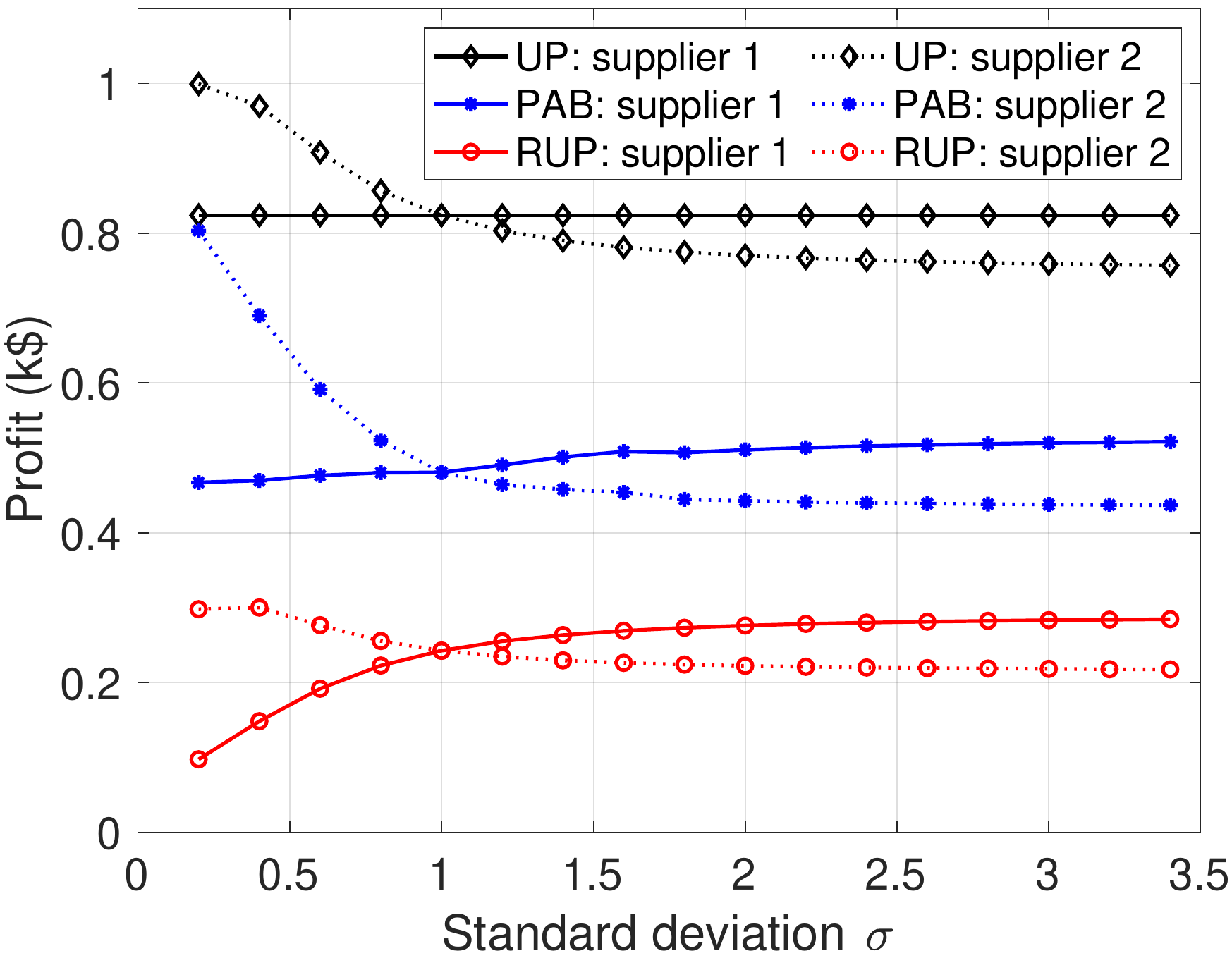}}}
	\vspace{-3ex}
	\caption{(a) Equilibrium price;  (b) Equilibrium profit. Both are with supplier 2's standard deviation of generations. } 
	\label{fig:syn}
	\vspace{-3ex}
\end{figure}

\begin{figure}[t]
	\centering
	\hspace{-6ex}
	\subfigure[]{
		
		\raisebox{-1mm}		{\includegraphics[width=1.65in]{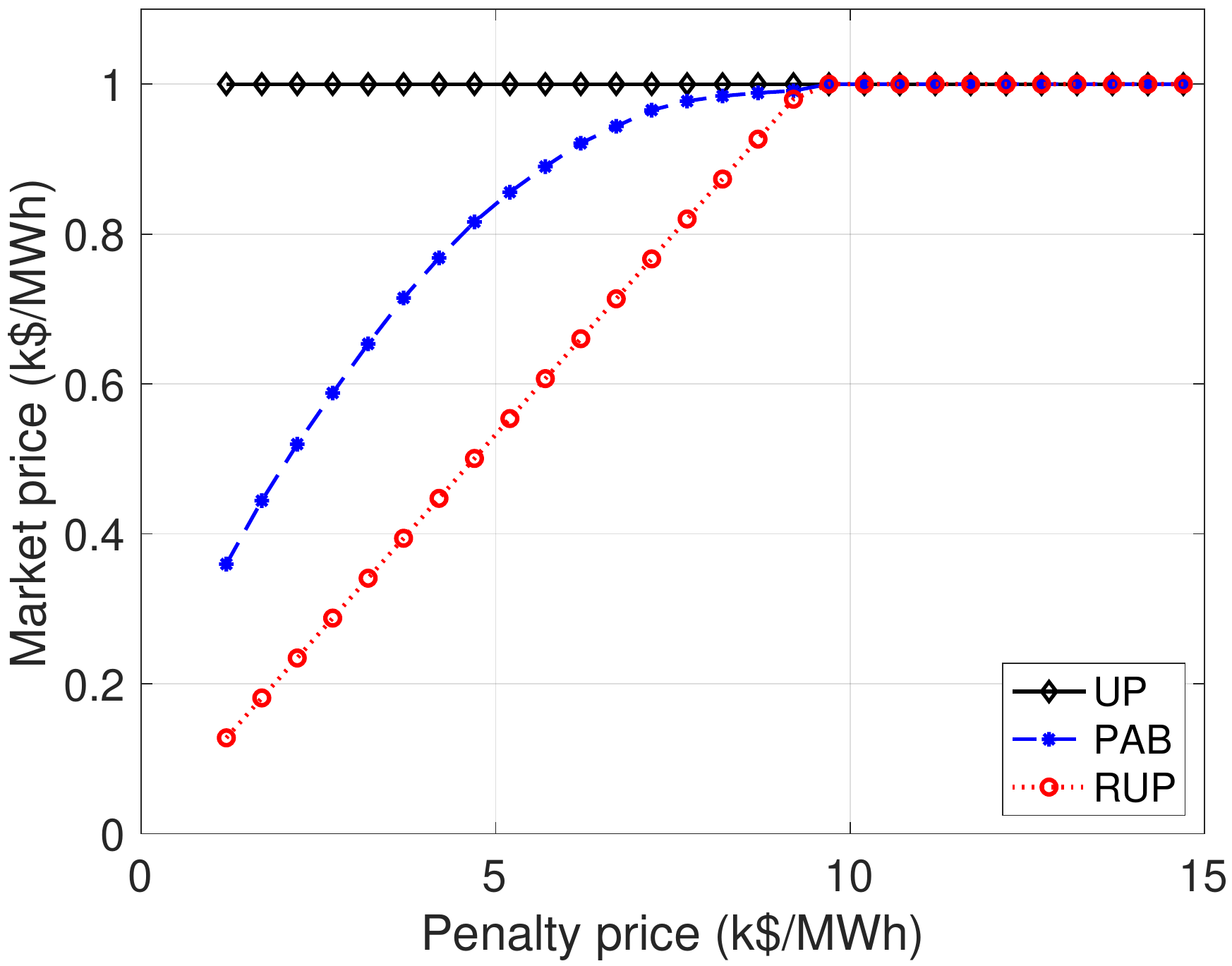}}}
	\hspace{-1ex}
	\subfigure[]{
		\raisebox{-1mm}		{\includegraphics[width=1.65in]{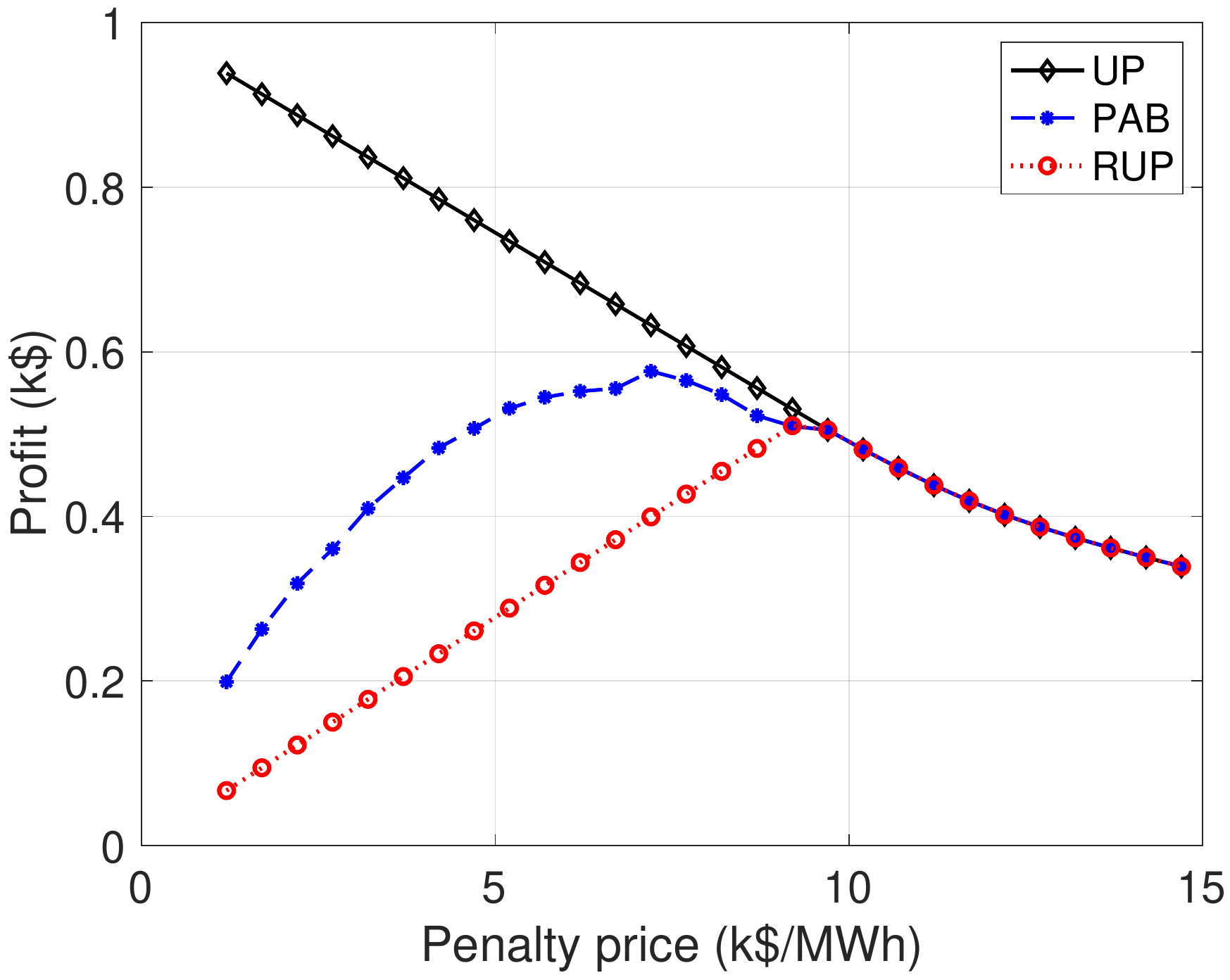}}}
	\vspace{-3ex}
	\caption{(a) Equilibrium price;  (b) Equilibrium profit, as a function of penalty price. } 
	\label{fig:real}
	\vspace{-3ex}
\end{figure}

 \subsection{Synthetic data}
 \subsubsection{Setup}
 For both suppliers, we assume their generations follow normal distributions with a mean value of 1.5MW, which are truncated between [0,3] MW. We fix supplier 1's standard deviation (std) at 1MW and vary supplier 2's. We set the real-time penalty price at $\lambda=1.5$k\$/MWh and assume two suppliers equally share the demand under \textbf{UP}.

  \subsubsection{Results}
  We show that \textbf{UP} can lead to the highest prices and profits for suppliers. Also, Under both  \textbf{PAB} and \textbf{RUP}, one supplier's increased generation uncertainty will reduce its own profit and increase the other supplier's profit.
  
  In Figures \ref{fig:syn}, we show equilibrium market prices in (a) and profits in (b) by varying supplier 2's std of generations.  \textbf{UP} will always lead to price cap $\bar{p}$ (black curve)  in Figure \ref{fig:syn}(a) and highest profits (black curve)  in Figure \ref{fig:syn}(b). As we assume that two suppliers equally share the demand, supplier 1's profit  (black solid curve) remains constant while supplier 2's profit  (black dotted curve) decreases as supplier 2's generation uncertainty increases.  Under \textbf{PAB}, in  Figure \ref{fig:syn}(a),  supplier 1's equilibrium bidding price (blue solid curve) increases while supplier 2's price (blue dotted curve)  decreases. The profits in Figure \ref{fig:syn}(b) show the same trend. The increased uncertainty of supplier 2 gives advantages to supplier 1 in market competition. Under \textbf{RUP}, in  Figure \ref{fig:syn}(a),  the uniform market price increases as supplier 2's generation uncertainty increases. The reason is that supplier 2 will bid more conservatively if its generation uncertainty increases and the operator needs to set a higher price to meet the demand. Accordingly, supplier 1's profit (red solid curve) will increase while supplier 2's profit (red dotted curve) will decrease as shown in Figure \ref{fig:syn}(b).

  \subsection{Real data}

   \subsubsection{Setup} We use the historical data of solar energy in Hong Kong from the year 1993 to the year 2012 \cite{zhao2019storage} to establish renewable-energy probability distribution. Specifically, for hour 4pm in July, we use  20-year historical data at this hour of this month to establish the empirical CDF of suppliers' renewable generations and approximate the continuous CDF \cite{nelson2013foundations}.   We focus on two homogeneous suppliers and vary the real-time penalty prices. We simulate the mixed price equilibrium by discretizing the price.  We assume two suppliers equally share the demand under \textbf{UP}.
  
  \subsubsection{Results}
  
\textbf{UP} will still lead to the highest prices and profits of suppliers, which is not beneficial to consumers, while   \textbf{RUP} will give the lowest prices and profits.  We also find that a higher real-time penalty price can both increase the market price and even increase a supplier's profit under \textbf{PAB} and \textbf{RUP}.

In Figures \ref{fig:real}, we show equilibrium market prices in (a) and profits in (b) by increasing the real-time penalty price $\lambda$.  \textbf{UP} will always lead to price cap $\bar{p}$ (black curve)  in Figure \ref{fig:real}(a) and highest profits (black curve)  in Figure \ref{fig:real}(b).  Under \textbf{PAB} and \textbf{RUP}, in  Figure \ref{fig:real}(a),  suppliers' equilibrium bidding prices (blue and red curves) increase, which can encourage suppliers to bid more quantities under the impact of higher penalty prices. This increased price actually improves suppliers'  profits(blue and red curves) in Figure \ref{fig:real}(b) when the penalty price is low. However, as the penalty price further increases, the penalty cost will dominate and suppliers' profits will decrease. By comparing between  \textbf{PAB} and \textbf{RUP},  the regulated \textbf{RUP} can achieve lower prices and profits, which are still always positive.

\section{Conclusion} \label{sec:con}
	
In this work, we analyze market equilibrium under different pricing mechanisms in a two-settlement  100\%-renewables  electricity market and provide insights into market design.  We establish game-theoretical models to compare equilibrium bidding strategies,  market prices, and profits between \textbf{UP} and \textbf{PAB}.  Without regulation, \textbf{UP} can induce the price cap while \textbf{PAB} can lead to lower market prices and profits. We present a new uniform-pricing mechanism \textbf{RUP} as a regulation benchmark, which can achieve even lower yet positive prices and profits compared with  \textbf{PAB}. Synthetic- and real-data simulations show that under \textbf{PAB} and \textbf{RUP} a higher uncertainty of renewables and a higher real-time shortage penalty price can both increase the market price by encouraging lower bidding quantities and even increase suppliers' profits. 

In future work, several assumptions can be generalized. For example, we will model elastic demand, provide a comprehensive analysis of real-time deviation prices,  and consider correlations between suppliers' random generations.

% \begin{acks}
% To Robert, for the bagels and explaining CMYK and color spaces.
% \end{acks}

\bibliographystyle{ACM-Reference-Format}
\bibliography{storage}

%%
%% If your work has an appendix, this is the place to put it.
\appendix

\begin{acks}
We would like to thank anonymous reviewers for their constructive
comments. 
This work has been supported  by ARPA-E Award No. DE-AR0001277.
\end{acks}

\section{Proof of Proposition \ref{prop:upq}}\label{app:prop1}

The Nash equilibrium given by \eqref{eq:upzeronash} can be easily proved based on Definition  \ref{def:pureprice}, where any unilateral deviation will not strictly increase a supplier's profit.

We now prove that  all the Nash equilibria satisfy $\pi^*=\bar{p}$ and $\sum_i q_i^*= D$. (i) suppose $\pi^*=0$ (i.e., $\sum_i q_i^*> D$) at the equilibrium. Any supplier bidding a positive quantity will get negative profits and it can always bid zero to get better off, which shows that $\pi^*=0$ is not the equilibrium.  (ii) Suppose $\sum_i q_i^*<D$ at the equilibrium. If there exists $j$ such that $q_j^*<y_j^*(\bar{p})$, this supplier $j$ can always increase its bid a bit to $q_j^*+\epsilon$ such that $\sum_{i\neq j} q_i^*+q_j^*+\epsilon<D$, which will increase its profit based on Lemma \ref{lem:quantity} and contradict the Nash equilibrium definition. If for any $i \in \mathcal{I}$, $q_i^*\geq y_i^*(\bar{p})$, i.e., $D>\sum_i q_i^*\geq \sum_i y_i^*(\bar{p})$, which contradicts 	$\sum_i y_i^*(\bar{p})> D$. \qed

\section{Generalization of bidding prices under uniform pricing}\label{app:generalize}

Although most current markets require suppliers to bid at marginal costs, they cannot fully capture the cost incurred by renewables' uncertainty in  real time. Therefore, we further generalize the setting and allow suppliers to bid any price instead of zero. One Nash equilibrium in the generalized case will coincide with Proposition \ref{prop:upq} together at price zero $p_i^*=0,\forall i$, where the clearing price reaches the price cap $\bar{p}$. This Nash equilibrium, however, may not be unique. We give another example of Nash equilibrium in Appendix B, where the clearing market price may not be at $\bar{p}$ but still be manipulated by marginal suppliers whose bidding price will be the clearing  price.

\subsection{Market-clearing mechanism} Each supplier bids the price $0\leq p_i\leq \bar{p}$ and quantity $0\leq q_i$. The commitment $\bm{x}^*$ and clearing market price $\pi^*$ are characterized  as follows.

First, the demand allocation $\bm{x}^*$ is based on merit order, which is equivalent to the following problem \textbf{MO}.\footnote{If some suppliers bid the same price, we assume that the system operator will allocate demand to these suppliers by a random  merit order.} For notation simplicity, we regard the system operator as supplier $i=0$ who bids at the price cap $p_0=\bar{p}$ and  quantity (shed load) $0\leq q_0\leq D$. We let the set $\mathcal{I}'=\mathcal{I}\bigcup \{0\}$.

\textbf{MO: Demand allocation based on merit order}
\begin{subequations}\label{eq:mo}
	\begin{align}
		\max_{\bm{x}}~ & \sum_{i\in \mathcal{I}'}(\bar{p}-p_i)\cdot x_i, \label{sg2:ob}\\
		\text{s.t.} ~~&\sum_{i\in \mathcal{I}'} x_i= D, \label{sg2:c1}\\
		~~&0\leq x_i \leq q_i,\forall i \in \mathcal{I}'. ~\label{sg2:c2}
	\end{align}
\end{subequations}

Then, we introduce how the cleared price $\pi^*$ is set. To begin with, we denote the set of suppliers who get positive committed quantity as   $I^\delta=\{i\in \mathcal{I}' |x_i^*>0 \}$. The supplier index with the highest price in  the set $I^\delta$ is $i^\alpha=\arg \max_{i\in \mathcal{I}^\delta }(p_i)$. The supplier index with the lowest price in the set $ \mathcal{I}'\setminus\mathcal{I}^\delta $ is $i^\beta=\arg \min_{i\in  \mathcal{I}' \setminus \mathcal{I}^\delta }(p_i)$. The cleared price $\pi^*$ is given by\footnote{We neglect the case of zero total bidding quantities from suppliers.}
\begin{align}
	\text{(i) If~}  x_{i^\alpha}^*<q_{i^\alpha}, \pi^*=p_{i^\alpha}.~	\text{(ii) If~} x_{i^\alpha}^*=q_{i^\alpha}, \pi^*=p_{i^\beta}. \label{eq:upnonzero}
\end{align}

\subsection{Game-theoretic model}
Any supplier $i\in\mathcal{I}$ decides both the bidding price $0\leq p_i\leq \bar{p}$ and quantity $0\leq q_i$ to maximize its profit.
\begin{align}
	R_i\left((p_i, q_i),\bm{p}_{-i},\bm{q}_{-i}\right)=&x_i^*(\bm{p},\bm{q}) \cdot  \pi^*(\bm{p},\bm{q}) \notag \\
	&- \mathbb{E}_{X_i}\left[\lambda\cdot  (x^*(\bm{p},\bm{q}) -X_i)^+\right],
\end{align}	
where  $\bm{x}^*$ is the solution to Problem \textbf{MO}  and $\pi^*$  is given by  \eqref{eq:upnonzero}. We have a definition of Nash equilibrium similar to Definition \ref{def:pureprice}.

\subsection{Nash equilibrium} The Nash equilibrium is affected by supply and demand. First, if there is supply shortage $\sum_i y_i^*({\bar{p}})\leq D$, any supplier $i$ just bids the quantity at $ y_i^*({\bar{p}})$, which leads to the price cap $\bar{p}$. The bidding price will not matter. Then, if $\sum_i y_i^*({\bar{p}})> D$, one Nash equilibrium will coincide with Proposition \ref{prop:upq}.

\begin{prop}[Nash equilibrium] \label{prop:uppq1}	If ~$\sum_{i \in \mathcal{I}} y_i^*(\bar{p})> D$,   we have one Nash equilibrium characterized by the conditions: $~p_i^*= 0,~\forall i\in \mathcal{I}$ and \eqref{eq:upzeronash}, 
	where the market clearing price is $\pi^{*}=\bar{p}$.
\end{prop}

\textbf{Proof:} We have the above proposition proved based on the Nash equilibrium definition. Suppose that any supplier $i$ deviates from the equilibrium strategy to $(p_i',q_i')$ unilaterally.
\begin{itemize}
    \item If $p_i'\geq 0$ and $q_i'+\sum_{k\in \mathcal{I}/i} q_k^*\leq D$,  the clearing price is still $\bar{p}$ but supplier $i$'s allocated demand $x_i'\leq q_i^*\leq y_i^*(\bar{p})$. Thus, the profit of supplier $i$ will not increase.
    \item If $p_i'\geq 0$ and $q_i'+\sum_{k\in \mathcal{I}/i} q_k^*>D$, the clearing price is now dropped to price $p_i'\leq \bar{p}$. Supplier $i$'s new allocated demand remains unchanged, i.e., $x_i'=q_i^*$ since its bidding price $p_i'\geq 0$ 
    Thus, the profit of supplier $i$ will not increase.\qed
\end{itemize}

At the equilibrium in Proposition \ref{prop:uppq1}, suppliers tend to bid low prices so as to be scheduled in the day-ahead market, and withhold bidding quantities to achieve high prices. This Nash equilibrium, however, may be not unique. We give another example next, where the clearing market price may not be at $\bar{p}$ but still be manipulated by marginal suppliers whose bidding price will determine  the clearing  price.

%In summary, under \textbf{UP}, suppliers can easily have the market power to incur high prices. Since  suppliers have zero marginal costs with uncertain generations, it can be hard for the market monitor to regulate their behaviors  like what it does to conventional controllable generators. Next, we discuss the results under \textbf{PAB}.

\subsection{Another example of Nash equilibrium}
We give another example of Nash equilibrium, where the clearing market price may not be at $\bar{p}$ but still may be manipulated by suppliers who bid high prices. 

\begin{prop}[Nash equilibrium] \label{prop:uppq2}	We have one Nash equilibrium characterized by the following conditions, where we define a new subset of suppliers  $\mathcal{I}^\gamma$ and let $j$ be the supplier index with the lowest price in the set $ \mathcal{I}\setminus\mathcal{I}^\gamma $, i.e.,  $j=\arg \min_{i\in  \mathcal{I} \setminus \mathcal{I}^\gamma }(p_i)$.
\begin{subequations}
	\begin{align}
		&\text{(i)} ~p_i^*= 0,~\forall i\in \mathcal{I}^\gamma,\\
		&\text{(ii)} ~ \sum_{i \in \mathcal{I}^\gamma } q_i^*= D,\\
		&\text{(iii)} ~ 0\leq  q_i^*\leq  y_i^*({p}_j), \forall i\in \mathcal{I}^\gamma, \\
		&\text{(iv)} ~   q_i^*\leq  q_j^*, \forall i\in \mathcal{I}^\gamma, 
	\end{align}
 \end{subequations}
	where the market clearing price is $\pi^{*}={p}^j$.
\end{prop}

\textbf{Proof:} We have the above proposition proved based on the Nash equilibrium definition. We discuss suppliers in sets $\mathcal{I}^\gamma$ and $\mathcal{I}\setminus \mathcal{I}^\gamma$, respectively.

First, suppose that supplier $i\in \mathcal{I}^\gamma$ deviates from the equilibrium strategy to $(p_i',q_i')$, unilaterally.
\begin{itemize}
    \item  $p_i'\geq 0$ and  $q_i'+\sum_{k\in \mathcal{I}^\gamma/i} q_k^*\leq D$: 
    \begin{itemize}
        \item If  $p_i'\leq p_j$, the clearing price is still $p_j$ since $q_i^*\leq q_j^*$. The supplier $i$'s allocated demand $x_i'\leq q_i^*\leq y_i^*(p_j)$. Thus, the profit of supplier $i$ will not increase.
        \item If $p_i'> p_j$, since $q_i^*\leq q_j^*$, the original allocated demand to supplier $i$ will be taken by supplier $j$. The new allocated demand is $x_i'=0$, which will decrease supplier $i$'s profit.
        
    \end{itemize}

    \item $p_i'\geq 0$ and   $q_i'+\sum_{k\in \mathcal{I}/i} q_k*>D$:
    \begin{itemize}
        \item If  $p_i'\leq p_j$, the clearing price is  $p_i'\leq p_j$. Since $p_i'\geq 0$, the allocated demand to supplier $i$ will not change, i.e., $x_i'= q_i^*$. Thus, the profit of supplier $i$ will not increase.
        \item If $p_i'> p_j$, since $q_i^*\leq q_j^*$, the original allocated demand to supplier $i$ will be taken by supplier $j$. The new allocated demand is $x_i'=0$, which will decrease supplier $i$'s profit.
        
    \end{itemize}

\end{itemize}

Second, suppose that supplier $i\in \mathcal{I}\setminus \mathcal{I}^\gamma$ deviates from the equilibrium strategy to $(p_i',q_i')$, unilaterally. Note that these suppliers get zero demand and zero profit at the equilibrium. 
\begin{itemize}
    \item  $p_i'>0$: supplier $i$ still cannot be allocated to demand and the profit will not change.

    \item $p_i'=0$ and  $q_i'=0$: the profit is still zero. 
    \item $p_i'=0$ and  $q_i'>0$: the market clearing price is dropped to $\pi^*=0$. The profit of supplier $i$ is no greater than zero.

\end{itemize}

Overall, we discussed all the cases where any supplier will not deviate unilaterally. We have Proposition \ref{prop:uppq2} proved. \qed

In Proposition \ref{prop:uppq2}, supplier $j$' bidding price will set the market price, which still can be very high.  Proposition  \ref{prop:uppq1} is a special case of Proposition  \ref{prop:uppq2} if in Proposition  \ref{prop:uppq2}, we regard the system operator as supplier $i=0$ and replace the set $\mathcal{I}$ by $\mathcal{I}'$.

\section{Proof of Proposition \ref{prop:pricecomp}} \label{app:proof3}
 We only need to prove $\pi^{\text{PAB}}\geq \pi^{\text{RUP}}>0$ when$ \sum_{i \in \mathcal{I} } y_i^*({\bar{p}})>D$. 
 
 (i) $\pi^{\text{PAB}}\geq \pi^{\text{RUP}}$: Proposition 4 in \cite{zhao2019storage} shows that the common lower support $l$ of  mixed price equilibrium of two suppliers satisfies  
$D \leq \sum_iy_i^*(l)$. Thus, we have $l\geq \pi^{\text{RUP}}$ since $D = \sum_iy_i^*(\pi^{\text{RUP}})$, which shows $\pi^{\text{PAB}}\geq \pi^{\text{RUP}}$. 

(ii) $\pi^{\text{RUP}}>0$: Under the mechanism \textbf{RUP}, if $0<\pi^*\leq \lambda$, the profit of supplier $i$ is 
$$\lambda \int_0^{F_{i}^{-1}\left(\frac{\pi^*}{\lambda}\right)} x f_i(x) d x>0.$$
Note that  $\pi^*> \lambda$ will not happen  as $Q(p)$ will remain constant when $p\geq \lambda$. Besides, since $D>0$, $\pi^*=0$ is impossible. Notably, $\pi^{\text{RUP}}>0$ will hold for oligopoly markets.
\qed
%\subsection{Part One}
%
%Lorem ipsum dolor sit amet, consectetur adipiscing elit. Morbi
%malesuada, quam in pulvinar varius, metus nunc fermentum urna, id
%sollicitudin purus odio sit amet enim. Aliquam ullamcorper eu ipsum
%vel mollis. Curabitur quis dictum nisl. Phasellus vel semper risus, et
%lacinia dolor. Integer ultricies commodo sem nec semper.
%
%\subsection{Part Two}
%
%Etiam commodo feugiat nisl pulvinar pellentesque. Etiam auctor sodales
%ligula, non varius nibh pulvinar semper. Suspendisse nec lectus non
%ipsum convallis congue hendrerit vitae sapien. Donec at laoreet
%eros. Vivamus non purus placerat, scelerisque diam eu, cursus
%ante. Etiam aliquam tortor auctor efficitur mattis.
%
%\section{Online Resources}
%
%Nam id fermentum dui. Suspendisse sagittis tortor a nulla mollis, in
%pulvinar ex pretium. Sed interdum orci quis metus euismod, et sagittis
%enim maximus. Vestibulum gravida massa ut felis suscipit
%congue. Quisque mattis elit a risus ultrices commodo venenatis eget
%dui. Etiam sagittis eleifend elementum.
%
%Nam interdum magna at lectus dignissim, ac dignissim lorem
%rhoncus. Maecenas eu arcu ac neque placerat aliquam. Nunc pulvinar
%massa et mattis lacinia.

\end{document}